\documentstyle[preprint,revtex]{aps} \voffset=-0.5in \hoffset=-0.25in
\begin{document} \draft \begin{title} Canonical quantization of $(2+1)$
dimensional QED with topological mass term \end{title} \author{Kurt
Haller and Edwin Lim-Lombridas} \begin{instit} Department of Physics,
University of Connecticut, Storrs, Ct 06269 \end{instit}
\begin{abstract} We discuss the canonical quantization of Quantum
Electrodynamics in $2+1$ dimensions, with a Chern-Simons topological
mass term and gauge-covariant coupling to a Dirac spinor field. A
gauge-fixing term is used which generates a canonical momentum for
$A_0$, so that there are no primary constraints on operator-valued
fields. Gauss's Law and the gauge condition, $A_0=0$, are implemented by
embedding the formulation in an appropriate physical subspace, in which
state vectors remain naturally, in the course of time evolution. The
photon propagator is derived from the canonical theory. The electric and
magnetic fields are separated into parts that reflect the presence of
massive photons, and other parts that are rigidly attached to charged
fermions and do not consist of any observable, propagating particle
excitations. The effect of rotations on charged particle states is
analyzed, and the relation between the canonical and the Belinfante
``symmetric'' angular momentum is discussed. It is shown that the
rotation operator can be consistently formulated so that charged
particles behave like fermions, and do not acquire any arbitrary phases
during rotations, even when they are dressed in the electromagnetic
fields required for them to obey Gauss's Law. \end{abstract} % \vspace{2.5in} % \noindent
% Preprint No.~UCONN-92-1, March, 1992\\ % Bitnet Address:
KHALLER@UCONNVM\\ % Internet Address: el@main.phys.uconn.edu \pacs{}
\narrowtext

\section{INTRODUCTION}

Gauge theories in $2+1$ dimensions have attracted considerable
attention, partly because of the suggestion that the quantum Hall
effect\cite{qhe} and high $T_c$ superconductivity\cite{rbletc} are
planar phenomena which such gauge theories might describe correctly.
Investigations of these gauge theories are also motivated by their
intrinsic theoretical interest.  One feature that is of interest in
$2+1$ dimensional Quantum Electrodynamics (QED) is the fact that a
gauge-invariant Chern-Simons (CS) term in the Lagrangian generates a
so-called topological mass for the single observable excitation mode of
the gauge field.\cite{djt}  The presence of the CS term is of further
interest because ``pure'' CS theory---i.e., the theory in which the CS
term is the only kinetic energy term in which the gauge field appears in
the Lagrangian---exhibits anyonic behavior.\cite{ch,s,ms,gvd}  Some
authors have suggested that when a Maxwell kinetic energy term is added
to pure CS theory, yielding $2+1$ dimensional QED with a CS topological
mass term, anyonic statistics should also
result.\cite{luscher,gwsandsodano,gwsandwije}

In this work, we will undertake the quantization of QED in $2+1$
dimensions with a topological mass term, interacting with a charged
fermion field which obeys the $2+1$ dimensional Dirac Equation. We will
work in the temporal gauge, and will apply a method of dealing with the
gauge field constraints (i.e., Gauss's Law and the gauge condition),
which one of us (KH) has used extensively in previous
work.\cite{temp1,temp2}  In this method, the constraints are not imposed
at the operator level, and a gauge-fixing term is chosen which generates
a momentum conjugate to $A_0$, the timelike component of the gauge
field. The primary constraint, that the momentum conjugate to $A_0$
vanishes, is thereby avoided, and the equal time commutation rules
(ETCR) are completely canonical. Because the theory is free of all
primary operator-valued constraints, no special procedures (such as
Dirac's)\cite{dirac} need to be used in its quantization; there are,
furthermore, no operator-ordering problems stemming from the failure of
constrained operators to commute. The gauge condition and Gauss's Law
are implemented in this procedure by selecting a physical subspace
within which all time evolution naturally takes place. Photon ghosts are
used to represent those gauge field components which mediate
interactions among charges and currents, but which do not have any
observable particle-like excitations. In this theory, these fields
include not only a cylindrically radial electric field, which is the
$2+1$ dimensional analog of the Coulomb field, but also a magnetic field
perpendicular to the spatial plane. Both these fields originate from the
same charge density sources.

In Sec.~II of this paper, we will apply the machinery of canonical
quantization to the Lagrangian for this model; we will also show how to
expand the space-dependent operator-valued fields in terms of plane wave
excitations for electrons and a single variety of propagating massive
photons, and for two varieties of gauge ghosts. In Sec.~III, we will
implement the constraints, and demonstrate that, once imposed, they
apply time-independently. In this section, we also will transform to a
representation in which the ``ghost'' degrees of freedom no longer
participate in the time evolution of observable states. When such a
transformation is carried out, non-local interactions describe the
long-range effects mediated by ghost excitations in the original
representation. In $3+1$ dimensional QED, this non-local interaction is
the familiar Coulomb interaction.\cite{temp1} In the model under
investigation here, the corresponding non-local interaction includes not
only the $2+1$ dimensional analog of the Coulomb interaction, but also a
further non-local interaction between charge and current densities; this
interaction reflects the existence of a magnetic field, originating from
a charged source, which has no observable particle-like excitations,
resembling, in that way, the Coulomb field in ordinary QED. In Sec.~IV,
we develop the perturbative theory for this model; we obtain the
propagator for the gauge field, and use it to calculate the leading
order of the S-matrix element for electron-electron scattering. In
Sec.~V, we will expand the gauge and spinor fields in terms of
excitations with definite angular momentum, and obtain an expression for
the total canonical (Noether) angular momentum operator. We will show
that the canonical angular momentum is invariant to time-independent
gauge transformations, which are the most general gauge transformations
that the canonical theory in the temporal gauge allows. We will also
exhibit other properties of the canonical angular momentum that make its
use as the rotation operator appropriate. We will apply the rotation
operator to gauge-invariant charged states, which obey Gauss's Law, and
we will show that, in contrast to pure CS theory, these charged states
only change sign in $2\pi$ rotations, and do not acquire arbitrary
``anyonic'' phases. Lastly, in this section we will discuss problems
that arise when the Belinfante ``symmetric'' form of the angular
momentum is used in the rotation operator.

\section{FORMULATION OF THE THEORY}

The Lagrangian for QED in $2+1$ dimensions with a topological mass term
is given by \begin{eqnarray} {\cal{L}}&=& -\frac{1}{4}F_{ln}F_{ln} +
\frac{1}{2}F_{l0}F_{l0} + j_{l}A_{l} - j_{0}A_{0} - \partial_{0}A_{0}G
\nonumber \\ &+& \frac{1}{4}m\epsilon_{ln}\left\{F_{ln} A_{0} - 2F_{n0}
A_{l} \right\} + \bar{\psi}(i\gamma^\mu \partial_\mu - M)\psi,
\label{eq:lagrangian2} \end{eqnarray} where $F_{ln} = \partial_nA_l -
\partial_lA_n$ and $F_{l0} = \partial_lA_0 + \partial_0A_l$; vector
components labeled with latin subscripts ($j_l$ and $A_l$) denote the
spatial components of the contravariant quantities ($j^\lambda$ and
$A^\lambda$, respectively); the latin subscripts are summed over two
spatial directions, and $\epsilon_{ln}$ is a completely antisymmetric
second rank tensor. The $\gamma$ matrices for the Dirac field will be
represented in terms of Pauli spin matrices as $\gamma^0 = -\sigma_3$,
$\gamma^1 = i\sigma_2$, and $\gamma^2 = -i\sigma_1$; the spinor currents
are $j^\lambda= e\bar{\psi}\gamma^\lambda\psi$.

Eq.~(\ref{eq:lagrangian2}) contains the CS term, as well as the gauge-
fixing term, $-G\partial_0A_0$, which enables us to implement the gauge
choice $A_0=0$, but which still avoids the primary constraint that
causes the momentum conjugate to $A_0$ to vanish as an operator
identity. Eq.~(\ref{eq:lagrangian2}) leads to the Euler-Lagrange
Equations \begin{equation} \partial_l F_{ln} + \partial_0 F_{n0} - j_n +
m\epsilon_{nl}F_{l0}=0, \label{eq:eqofmotioncs1} \end{equation}
\begin{equation} \partial_l F_{l0} + j_0 -
\frac{1}{2}m\epsilon_{ln}F_{ln} = \partial_0 G, \label{eq:eqofmotioncs2}
\end{equation} \begin{equation} \partial_0 A_0 = 0,
\label{eq:eqofmotioncs3} \end{equation} and \begin{equation} (M -
i\gamma^\mu D_\mu)\psi = 0, \label{eq:diraceq} \end{equation} where
$D_\mu$ is the gauge-covariant derivative $D_{\mu} = \partial_{\mu} +
ieA_{\mu}$. Eq.~(\ref{eq:eqofmotioncs1}) is the analog of the Maxwell-
Ampere Equation for QED in $2+1$ dimensions with a topological mass
term; in the limit $m \rightarrow 0$, it is just the Maxwell-Ampere
Equation in $2+1$ dimensional massless QED. When $\partial_l F_{ln}$ and
$\partial_0F_{n0}$ are eliminated, it becomes the corresponding equation
in pure CS Theory. Eq.~(\ref{eq:eqofmotioncs2}) would be the analog of
Gauss's Law in this theory if its right-hand side were 0 instead of
$\partial_0G$. In that case, the $m \rightarrow 0$ limit again leads to
Gauss's Law for $2+1$ dimensional massless QED; and the elimination of
$\partial_l F_{l0}$ leads to $j_0 = (1/2)m\epsilon_{ln}F_{ln}$, the form
that Gauss's Law takes in pure CS theory. The inclusion of $\partial_0G$
introduces the time dependence that makes Eq.~(\ref{eq:eqofmotioncs2})
an equation of motion instead of a time-independent operator-valued
constraint. An additional equation, important for the implementation of
Gauss's Law, is \begin{equation} \partial_0 \partial_0 G = 0;
\label{eq:DoDoG2} \end{equation} it is obtained by differentiating
Eq.~(\ref{eq:eqofmotioncs1}) by $\partial_n$ and
(\ref{eq:eqofmotioncs2}) by $\partial_0$, and adding the resulting two
equations. In Sec. III, we will show how Eqs.~(\ref{eq:eqofmotioncs3})
and (\ref{eq:DoDoG2}) enable us to chose a physical subspace within
which the time evolution of state vectors naturally is contained, and in
which the gauge condition, $A_0=0$, and Gauss's Law hold.

The momenta conjugate to the fields are \begin{equation} \Pi_n =
\frac{\partial {\cal{L}}}{\partial(\partial_0 A_{n})} = F_{n0} +
\frac{1}{2}m\epsilon_{nl} A_l, \label{eq:pin2} \end{equation}
\begin{equation} \Pi_0 = \frac{\partial {\cal{L}}}{\partial(\partial_0
A_{0})} = -G, \label{eq:pi02} \end{equation} \begin{equation} \Pi_\psi =
\frac{\partial {\cal{L}}}{\partial(\partial_0 \psi)} = i\psi^{\dagger};
\label{eq:pipsi} \end{equation} and the Hamiltonian density is ${\cal H}
= {\cal H}_0 + {\cal H}_I$ where \begin{eqnarray}
{\cal{H}}_0&=&\frac{1}{2}\Pi_n \Pi_n + \frac{1}{4}F_{ln}F_{ln} + A_0
\partial_n \Pi_n + \frac{1}{8}m^2 A_l A_l\nonumber \\
&+&\frac{1}{2}m\epsilon_{ln}A_l \Pi_n -
\frac{1}{4}m\epsilon_{ln}F_{ln}A_0 +{\cal{H}}_{e\bar{e}}
\label{eq:hdensity2a} \end{eqnarray} and \begin{equation} {\cal H}_I = -
j_n A_n + j_0 A_0; \end{equation} ${\cal H}_{e\bar{e}}$ is the
Hamiltonian density for the electron field and is given by ${\cal
H}_{e\bar{e}} = \psi^\dagger(\gamma^0M -
i\gamma^0\gamma^n\partial_n)\psi$. Since there are no primary
constraints in this formulation of the theory, the equal time
commutation (or anticommutation) rules are completely canonical, and are
given by \begin{equation} {[}A_0({\bf x}), G({\bf y}){]} = -i\delta({\bf
x - y}), \label{eq:equaltimecomm1} \end{equation} \begin{equation}
{[}A_l({\bf x}), \Pi_n({\bf y}){]} = i\delta_{l,n}\delta({\bf x - y}),
\label{eq:equaltimecomm2} \end{equation} \begin{equation} \{\psi_\alpha
({\bf x}), \psi^\dagger_\beta({\bf y})\} =
\delta_{\alpha,\beta}\delta({\bf x - y}). \label{eq:equaltimeanticomm}
\end{equation}

In order to arrive at a Fock Space in which time evolution of state
vectors takes place, we represent the space-time fields in terms of
creation and annihilation operators for the appropriate gauge field
excitations. Since there is only one, massive, observable gauge field
excitation, we surmise that the transverse mode of $A_l$ can be
represented by $A_l^T = e^{i\eta}\sum_{\bf
k}{(\epsilon_{ln}k_{n}/k)({2\omega})^{-1/2}a({\bf k}) + \mbox{\rm
h.a.}}$, and the transverse mode of $\Pi_l$ by $\Pi^T_l =
e^{i\eta^\prime}\sum_{\bf k}{\epsilon_{ln}}$ $(k_{ln}/k)$
$(\omega/2)^{1/2}a({\bf k}) + \mbox{\rm h.a.}$, where $\eta$ and
$\eta^\prime$ are arbitrary phase factors; $a({\bf k})$ and
$a^\dagger({\bf k})$ obey ${[}a({\bf k}), a^{\dagger}({\bf q}){]} =
\delta_{\bf k,q}$ and annihilate and create, respectively, photons of
momentum ${\bf k}$ and energy $\omega$, where $\omega = \sqrt{m^2 +
k^2}$. Since this single variety of gauge field excitation is
insufficient to represent all the commutation rules given in
Eqs.~(\ref{eq:equaltimecomm1})--(\ref{eq:equaltimeanticomm}), we must
use additional excitations to complete the momentum space representation
of all fields in this theory. Proceeding under the assumption that there
is indeed only one observable excitation mode in this gauge field, we
make the {\it ansatz} that ghost modes alone will suffice to complete
the desired Fock Space. In fact, the identical ghost spectrum used in
previous work is well suited for this purpose.\cite{temp1} These ghost
modes come in two varieties, $a_Q({\bf k})$ and $a_R({\bf k})$, and
their hermitian adjoints are  $a_Q^\star({\bf k})$ and $a_R^\star({\bf
k})$, respectively. They obey the ETCR \begin{equation} {[}a_Q({\bf k}),
a_R^{\star}({\bf q}){]}={[}a_R({\bf k}), a_Q^{\star}({\bf q}){]} =
\delta_{\bf k,q}; \label{eq:commutatorofexcitation1} \end{equation} and
\begin{equation} {[}a_Q({\bf k}), a_Q^{\star}({\bf q}){]}={[}a_R({\bf
k}), a_R^{\star}({\bf q}){]}=0. \label{eq:commutatorofexcitation2}
\end{equation} The norm of any state with a single variety of these
ghosts, e.g., the norm of the one particle state, $||a_Q^\star({\bf
k})N\rangle|$ $=$ $\langle Na_Q({\bf k})|a_Q^\star({\bf k})N\rangle$,
vanishes identically, where $|N\rangle$ represents a state in which any
assemblage of creation operators for observable particle modes act on
the vacuum state $|0\rangle$ annihilated by all annihilation operators.
The unit operator in the one particle ghost (opg) sector is given by
\begin{equation} [1]_{\mbox{opg}} = \sum_{\bf k}{a_Q^\star({\bf
k})|0\rangle\langle 0|a_R({\bf k}) + a_R^\star({\bf k})|0\rangle\langle
0|a_Q({\bf k})}. \end{equation} The use of ghosts is appropriate and
necessary for components of gauge fields which have non-vanishing
commutators with each other, but which do not exhibit any observable,
propagating excitations. The {\it ansatz} that the further excitations
required to represent a gauge field are ghosts, therefore tests the
surmise that the single massive photon is the only observable mode it
has. To find a useful representation of the longitudinal fields, as well
as of $A_0$ and $G$, in terms of ghost excitations, we undertake to find
a representation that will lead to $a_Q({\bf k})$ and $a_Q^\star({\bf
k})$ operators that are time-independent in the interaction-free case,
i.e., a representation for which $[H_0, a_Q({\bf k})] =0$ and $[H_0,
a_Q^\star({\bf k})]=0$. To find such a representation, we use a
conveniently chosen but arbitrary set of operators, $\varphi_i({\bf k})$
and $\varphi_i^\star({\bf k})$, obeying commutations rules
$[\varphi_i({\bf k}), \varphi_j^\star({\bf q})] = C_{i,j}\delta_{\bf
k,q}$, where $C_{i,j}$ is a symmetric matrix. We also represent
$a_Q({\bf k})$ and $a_R({\bf k})$, and their adjoints, as
\begin{eqnarray} a_Q({\bf k}) = \sum_{i}{Q_i\varphi_i({\bf k})}, &&
a_R({\bf k}) = \sum_{i}{R_i\varphi_i({\bf k})},\nonumber \\
a_Q^\star({\bf k}) = \sum_{i}{Q_i^*\varphi_i^\star({\bf k})}, &&
a_R^\star({\bf k}) = \sum_{i}{R_i^*\varphi_i^\star({\bf k})},
\label{eq:varphi} \end{eqnarray} constraining the coefficients $Q_i$,
$R_i$ so that the commutation rules,
Eqs.~(\ref{eq:commutatorofexcitation1}) and
(\ref{eq:commutatorofexcitation2}), are satisfied. We then expand the
longitudinal components of $A_l$ and $\Pi_l$, as well as $A_0$ and $G$,
in terms of the $\varphi_i({\bf k})$ and $\varphi_i^\star({\bf k})$
excitations, again  using a simple {\it ansatz}. We then invert
Eq.~(\ref{eq:varphi}) and express the Hamiltonian $H_0 = \int{{\cal
H}_0({\bf x})d{\bf x}}$ in terms of the operators $a_Q({\bf k})$,
$a_R({\bf k})$, $a_Q^\star({\bf k})$ and $a_R^\star({\bf k})$, and the
coefficients $Q_i$, $R_i$, $Q_i^*$ and $R_i^*$. We adjust the
coefficients so that the ghost operators $a_R({\bf k})$ and
$a_R^\star({\bf k})$ do not appear in $H_0$, and therefore that $H_0$
commutes with $a_Q({\bf k})$ and $a_Q^\star({\bf k})$. These
calculations were carried out using an operator algebra manipulation
package in {\it Mathematica}\cite{math}. The fact that this program can
be successfully completed serves as confirmation
that only one observable, massive mode arises in this model. As a last
step, we unitarily transform all operators to eliminate combinations of
the form $a_Q({\bf k})a(-{\bf k})$, $a_Q^\star({\bf k})a({\bf k})$,
$a_Q({\bf k})a^\dagger({\bf k})$ and $a_Q^\star({\bf k})a^\dagger(-{\bf
k})$ from the transformed $H_0$. The resulting fields have the form
\widetext \begin{eqnarray} A_l ({\bf x}) &=& -
\frac{m}{\sqrt{2}}\sum_{{\bf k}}{\frac{k_l}{k \omega^{3/2}}\left[ a({\bf
k}) e^{i{\bf k \cdot x}} + a^{\dagger}({\bf k})e^{-i{\bf k \cdot
x}}\right]}\nonumber\\ &+& \frac{i}{\sqrt{2}}\sum_{{\bf
k},\,n=1,2}{\frac{\epsilon_{ln}k_n}{k \sqrt{\omega}}\left[ a({\bf k})
e^{i{\bf k \cdot x}} - a^{\dagger}({\bf k})e^{-i{\bf k \cdot
x}}\right]}\nonumber\\ &+&\frac{1}{16}\sum_{{\bf k}}{\frac{v^4 k_l}{k^3
\kappa_0^{5/2}}\left[a_R({\bf k}) e^{i{\bf k \cdot x}} +
a_R^{\star}({\bf k})e^{-i{\bf k \cdot x}}\right]}\nonumber\\ &-&
\frac{8\gamma}{\beta} \sum_{{\bf k}}{\frac{k \kappa_0^{3/2}k_l}{v^2
\omega^2}\left[a_Q({\bf k}) e^{i{\bf k \cdot x}} + a_Q^{\star}({\bf
k})e^{-i{\bf k \cdot x}}\right]}\nonumber\\ &+&8im \sum_{{\bf
k},\,n=1,2}{\frac{\epsilon_{ln}k_n k \kappa_0^{5/2}}{v^4
\omega^2}\left[a_Q({\bf k}) e^{i{\bf k \cdot x}} - a_Q^{\star}({\bf
k})e^{-i{\bf k \cdot x}}\right]}, \label{eq:Alcs} \end{eqnarray}
\begin{eqnarray} \Pi_l ({\bf x}) &=& \frac{1}{\sqrt{2}}\sum_{{\bf k},\,
n=1,2}{\frac{\epsilon_{ln}k_n\sqrt{\omega}}{k}\left\{1 -
\frac{m^2}{2\omega^2}\right\}\left[ a({\bf k}) e^{i{\bf k \cdot x}} +
a^{\dagger}({\bf k})e^{-i{\bf k \cdot x}}\right]}\nonumber\\
&+&\frac{im}{2^{3/2}}\sum_{{\bf k}}{\frac{k_l}{k \sqrt{\omega}}\left[
a({\bf k}) e^{i{\bf k \cdot x}} - a^{\dagger}({\bf k})e^{-i{\bf k \cdot
x}}\right]}\nonumber\\ &+& \frac{m}{32}\sum_{{\bf k},\,n=1,2}{\frac{v^4
\epsilon_{ln}k_n}{k^3 \kappa_0^{5/2}}\left[a_R({\bf k}) e^{i{\bf k \cdot
x}} + a_R^{\star}({\bf k})e^{-i{\bf k \cdot x}}\right]}\nonumber\\ &-
&\frac{4m\gamma}{\beta} \sum_{{\bf k},\,n=1,2}{\frac{\epsilon_{ln} k
\kappa_0^{3/2}k_n}{v^2 \omega^2}\left[a_Q({\bf k}) e^{i{\bf k \cdot x}}
+ a_Q^{\star}({\bf k})e^{-i{\bf k \cdot x}}\right]}\nonumber\\ &+& 4im^2
\sum_{{\bf k}}{\frac{k_l k \kappa_0^{5/2}}{v^4 \omega^2}\left\{1 -
\frac{2\omega^2}{m^2}\right\}\left[a_Q({\bf k}) e^{i{\bf k \cdot x}} -
a_Q^{\star}({\bf k})e^{-i{\bf k \cdot x}}\right]}, \label{eq:Plcs}
\end{eqnarray} \narrowtext \FL \begin{equation} A_0({\bf x}) = -4i
\gamma \sum_{{\bf k}}{\frac{k\sqrt{\kappa_0}}{v^2}\left[a_Q({\bf k})
e^{i{\bf k \cdot x}} - a_Q^{\star}({\bf k})e^{-i{\bf k \cdot
x}}\right]}, \label{eq:A0cs} \end{equation} and \FL \begin{eqnarray}
G({\bf x}) &=& \frac{1}{8\gamma} \sum_{{\bf
k}}{\frac{v^2}{k\sqrt{\kappa_0}} \left[a_R({\bf k}) e^{i{\bf k \cdot x}}
+ a_R^{\star}({\bf k})e^{-i{\bf k \cdot x}}\right]}\nonumber\\ &+&
\frac{16}{\beta} \sum_{{\bf k}}{\frac{k^3 \kappa_0^{7/2}}{v^4
\omega^2}\left[a_Q({\bf k}) e^{i{\bf k \cdot x}} + a_Q^{\star}({\bf
k})e^{-i{\bf k \cdot x}}\right]}, \label{eq:Gcs} \end{eqnarray}
\noindent where $\gamma$ and $\beta$ are arbitrary, real numerical
parameters, and where  $v^{2} = m^{2} + 8k^{2}$, $\kappa_0^2 = 4k^2 +
m^2$ and $\omega^{2} = m^{2} + k^{2}$. The electric and magnetic fields
are \widetext \begin{eqnarray} E_l({\bf x}) &=& -\sum_{{\bf
k},n=1,2}{\frac{\epsilon_{ln}k_n}{k}\sqrt{\frac{\omega}{2}}\left\{a({\bf
k})e^{i{\bf k \cdot x}} + a^\dagger({\bf k})e^{-i{\bf k \cdot
x}}\right\}}\nonumber\\ &-&\sum_{\bf
k}{\frac{imk_l}{k\sqrt{2\omega}}\left\{a({\bf k})e^{i{\bf k \cdot x}} -
a^\dagger({\bf k})e^{-i{\bf k \cdot x}}\right\}}\nonumber\\ &+&
\sum_{\bf k}{\frac{8ik_l k^3 \kappa_0^{5/2}}{v^4
\omega^4}\left\{a_Q({\bf k})e^{i{\bf k \cdot x}} - a_Q^\star({\bf
k})e^{-i{\bf k \cdot x}}\right\}} \label{eq:electricfield}
\end{eqnarray} \narrowtext \noindent and \FL \begin{eqnarray} B({\bf x})
&=& \sum_{\bf k}{\frac{k}{\sqrt{2\omega}}\left\{a({\bf k})e^{i{\bf k
\cdot x}} + a^\dagger({\bf k})e^{-i{\bf k \cdot x}}\right\}}\nonumber\\
&+& \sum_{\bf k}{\frac{8m k^3 \kappa_0^{5/2}}{v^4
\omega^2}\left\{a_Q({\bf k})e^{i{\bf k \cdot x}} + a_Q^\star({\bf
k})e^{-i{\bf k \cdot x}}\right\}} \label{eq:magneticfield}.
\end{eqnarray} When we substitute these expressions for $A_l$, $\Pi_l$,
and $A_0$ into Eq.~(\ref{eq:hdensity2a}), we obtain $H_0$ in the
following form: \begin{eqnarray} H_0 &=& \sum_{\bf k}{
\frac{\omega}{2}\left\{a({\bf k})a^{\dagger}({\bf k}) + a^{\dagger}({\bf
k})a({\bf k})\right\}}\nonumber\\ &+&\sum_{\bf k}{ \frac{32k^6
\kappa_0^5}{v^8 \omega^2}\left\{(1-i\frac{\gamma v^2 \omega^2}{k^2
\kappa_0^2})a_Q(-{\bf k})a_Q({\bf k})\right.}\nonumber\\ &+&\left.
2a_Q^{\star}({\bf k})a_Q({\bf k}) + (1+i\frac{\gamma v^2 \omega^2}{k^2
\kappa_0^2})a_Q^{\star}(-{\bf k})a_Q^{\star}({\bf k})\right\}\nonumber\\
&+& H_{e\bar{e}}, \label{eq:hfreegauge2} \end{eqnarray} where
$H_{e\bar{e}} = \int{d{\bf x}}\,{\cal H}_{e\bar{e}}$. We can construct a
Fock Space, $\{|h_i\rangle\}$, based on the perturbative vacuum,
$|0\rangle$, which is annihilated by all the annihilation operators,
$a({\bf k})$, $a_Q({\bf k})$ and $a_R({\bf k})$, as well as the electron
and positron operators $e({\bf k})$ and $
\bar{e}({\bf k})$.\cite{comment} We observe that, in this perturbative
Fock Space, all multiparticle states of observable particles that
include electrons, positrons, and the massive photons created by
$a^\dagger({\bf k})$, are eigenstates of $H_0$. States, in which a
single variety of ghost creation operator such as $a_Q^\star({\bf k})$
or $a_Q^\star({\bf k}_1)a_Q^\star({\bf k}_2)$, operates on one of these
multiparticle states $|N\rangle$, have zero norm; they have no
probability of being observed, and have vanishing expectation values of
$H_0$, momentum, as well as of all other observables for which they are
eigenstates. States, in which both varieties of ghost appear
simultaneously such as $a_Q^\star({\bf k}_1)a_R^\star({\bf
k}_2)|N\rangle$, are not interpretable, and their appearance in the
course of time evolution signals the danger of a catastrophic defect in
the theory.

Substitution of Eqs.~(\ref{eq:Alcs})--(\ref{eq:Gcs}) into $H_I$ leads to
an expression in which all gauge field excitations appear, including
creation and annihilation operators for both varieties of ghosts. In
Sec. III, we will show how implementation of the constraints prevents
the catastrophic appearance of state vectors in which both varieties of
ghosts coincide.

\section{IMPLEMENTING GAUSS'S LAW\\ AND GAUGE CONDITION}

The operator, ${\cal G}$, that expresses Gauss's Law in this model, is
\begin{equation} {\cal G} = \partial_nF_{n0} + j_0 -
\frac{1}{2}m\epsilon_{nl}F_{nl}, \end{equation} so that
Eq.~(\ref{eq:eqofmotioncs2}) is given as ${\cal G} = \partial_0G$;
${\cal G}$ can also be represented as \begin{equation} {\cal G} =
\partial_n\Pi_n + j_0 - \frac{1}{4}m\epsilon_{nl}F_{nl}. \label{eq:calG}
\end{equation} Substitution of Eqs.~(\ref{eq:Alcs})--(\ref{eq:Gcs}) into
Eq.~(\ref{eq:calG}) leads to \FL \begin{equation} {\cal G} = \sum_{\bf
k}{\frac{8k^{3}\kappa_0^{5/2}}{v^{4}}\left\{a_{Q}({\bf k})e^{i{\bf
k}\cdot{\bf x}}+a_{Q}^{\star}({\bf k})e^{-i{\bf k}\cdot{\bf x}}\right\}}
+ j_{0}({\bf x}). \label{eq:DoG2} \end{equation} We can write this as
\begin{equation} {\cal G}  = \sum_{\bf
k}{\frac{8k^{3}\kappa_0^{5/2}}{v^{4}}\left\{\Omega({\bf k})e^{i{\bf
k}\cdot{\bf x}}+\Omega^{\star}({\bf k})e^{-i{\bf k}\cdot{\bf
x}}\right\}} \label{eq:DoG2a} \end{equation} where $\Omega({\bf k})$ is
defined by \begin{equation} \Omega({\bf k}) = a_Q({\bf k}) +
\frac{v^4}{16k^3 \kappa_0^{5/2}}j_{0}({\bf k}). \label{eq:omega2}
\end{equation} Similarly, we can express $A_0$ as \FL \begin{equation}
A_0 ({\bf x}) = -4i \gamma \sum_{{\bf
k}}{\frac{k\sqrt{\kappa_0}}{v^2}\left[\Omega({\bf k}) e^{i{\bf k \cdot
x}} - \Omega^{\star}({\bf k})e^{-i{\bf k \cdot x}}\right]}.
\label{eq:A0csOmega} \end{equation}

We will now implement Gauss's Law and $A_0 = 0$ by defining a ``physical
subspace'' $\{|\nu\rangle\}$, of another Fock Space, in which all state
vectors $|\nu\rangle$ obey the condition \begin{equation} \Omega({\bf
k})|\nu\rangle = 0. \label{eq:constraint} \end{equation} For all state
vectors $|\nu\rangle$ and $|\nu^\prime\rangle$ in this physical subspace
$\{|\nu\rangle\}$, $\langle\nu^\prime|{\cal G}|\nu\rangle = 0$ and
$\langle\nu^\prime|A_0|\nu\rangle = 0$, so that both Gauss's Law and the
gauge condition hold. Moreover, a state vector initially in the physical
subspace will always remain entirely contained in it, as it develops
under time evolution. This follows from the fact that $[H,\Omega({\bf
k})]=0$, a relation which can be explicitly demonstrated, as well as
inferred from Eqs.~(\ref{eq:eqofmotioncs3}) and (\ref{eq:DoDoG2}). To
complete the Fock Space in which this physical subspace is embedded, we
note that there is a unitary transformation, $U = e^{iD}$, for which
$U^{-1}\Omega({\bf k})U = a_Q({\bf k})$, where $D$ is given by \widetext
\begin{equation} D = \frac{1}{2}\int{d{\bf x}d{\bf y}\,j_0({\bf
y})\left\{\xi(|{\bf x - y}|)\frac{\partial}{\partial x_l} A_l({\bf x}) +
\eta(|{\bf x - y}|)\epsilon_{ln}\frac{\partial}{\partial x_l}\Pi_n({\bf
x})\right\}} \label{eq:Dspace} \end{equation} \narrowtext \noindent and
where $\xi$ and $\eta$ are given by \begin{equation} \xi(|{\bf x - y}|)
= -i\sum_{\bf k}{\left\{\frac{1}{k^2} +
\frac{1}{\omega^2}\right\}e^{i{\bf k \cdot (x-y)}}}, \end{equation} and
\begin{equation} \eta(|{\bf x - y}|) = 2im\sum_{\bf k}{\frac{e^{i{\bf k
\cdot (x-y)}}}{k^2\omega^2}}. \end{equation} $D$ can also be expressed
as \begin{eqnarray} D &=& \sum_{\bf k}{\frac{v^4}{16 k^3
\kappa_0^{5/2}}\left\{a_{R}({\bf k})j_{0}(-{\bf k}) - a_{R}^{\star}({\bf
k})j_{0}({\bf k})\right\}}\nonumber\\ &-&\sum_{\bf k}{\frac{8\gamma
\kappa_0^{3/2}k}{\beta v^2 \omega^2}\left\{a_{Q}({\bf k})j_{0}(-{\bf k})
- a_{Q}^{\star}({\bf k})j_{0}({\bf k})\right\}}. \label{eq:D}
\end{eqnarray} We can use $U$ to construct a set of state vectors
$U|n\rangle$, for a subspace $\{|n\rangle\}$ of the previously
established Hilbert Space $\{|h_i\rangle\}$, such that the state vectors
$U|n\rangle$ constitute the physical subspace $\{|\nu\rangle\}$. The
required state vectors $|n\rangle$ are those in which $a_Q^\star({\bf
k})$ and $a^\dagger({\bf k})$, as well as electron and positron creation
operators, act on the perturbative vacuum state $|0\rangle$. But no
$a_R^\star({\bf k})$ operators may appear in the $|n\rangle$ for which
$U|n\rangle$ comprise the subspace $\{|\nu\rangle\}$, since for states,
$|h_r\rangle$, in which $a_R^\star({\bf k})$ operators act on the
observable multiparticle states $|N\rangle$, $a_Q({\bf
k})|h_r\rangle\neq 0$; and, therefore, $\Omega({\bf k})|\rho\rangle\neq
0$ for the states $|\rho\rangle=U|h_r\rangle$.

It is convenient to establish an entirely equivalent, alternative
formalism, in which all operators and states are unitarily transformed
by the unitary transformation $U$. Since all matrix elements and
eigenvalues are invariant to such a similarity transformation, we can
construct a map $\{|\nu\rangle\} \rightarrow \{|n\rangle\}$,
$\Omega({\bf k}) \rightarrow a_Q({\bf k})$, and, in general, for all
other operators ${\cal P}$, ${\cal P} \rightarrow \tilde{\cal P}$, where
$\tilde{\cal P} = U^{-1}{\cal P}U$. We may then use the transformed
representation as an equivalent formulation of the theory, in which
Gauss's Law and the gauge constraint, $A_0=0$, have been implemented. In
that case, it becomes of particular significance to find the transformed
Hamiltonian and the transformed field operators. Using the Baker-
Hausdorff-Campbell formula, we find that \begin{equation} \tilde{H} =
H_0 + \tilde{H}_{int} \end{equation} where $H_0$ is untransformed and
\widetext \begin{eqnarray} \tilde{H}_{int} &=& \frac{1}{4\pi}\int{d{\bf
x}d{\bf y}\,j_0({\bf x})j_0({\bf y})K_0(m|{\bf x - y}|)}\nonumber\\ &+&
\int{d{\bf x}d{\bf y}\,j_0({\bf x})\epsilon_{ln}j_l({\bf y})({x -
y})_n{\cal F}(m|{\bf x - y}|)}\nonumber \\ &+&\sum_{{\bf
k},\,l=1,2}{\frac{mk_{l}}{ \sqrt{2}k \omega^{3/2}}\left\{j_{l}({\bf
k})a^{\dagger}({\bf k}) + j_{l}(-{\bf k})a({\bf k})\right\}}\nonumber\\
&+& \sum_{{\bf k},\,l,n=1,2}{\frac{i\epsilon_{ln}k_{n}}{k\sqrt{2\omega}}
\left\{j_{l}({\bf k})a^{\dagger}({\bf k}) - j_{l}(-{\bf k})a({\bf
k})\right\}}\nonumber\\ &-&\sum_{\bf k}{\frac{8k^3 \kappa_0^{5/2}}{v^4
\omega^2}\left\{j_{0}({\bf k})a_{Q}^{\star}({\bf k}) + j_{0}(-{\bf
k})a_{Q}({\bf k})\right\}}\nonumber\\ &+&\sum_{{\bf
k},\,l,n=1,2}{\frac{8im k \kappa_0^{5/2} \epsilon_{ln}k_{n}}{v^4
\omega^2}\left\{j_{l}({\bf k})a_{Q}^{\star}({\bf k}) - j_{l}(-{\bf
k})a_{Q}({\bf k})\right\}}\nonumber\\ &+&\sum_{{\bf
k},\,l=1,2}{\frac{8\gamma k \kappa_0^{3/2} k_{l}}{\beta v^2
\omega^2}\left\{j_{l}({\bf k})a_{Q}^{\star}({\bf k}) + j_{l}(-{\bf
k})a_{Q}({\bf k})\right\}} \label{eq:HItilde} \end{eqnarray} \narrowtext
\noindent where $K_0(x)$ is a modified Bessel function, and
\begin{equation} {\cal F}(mQ) \equiv
\frac{m^2}{Q}\int_0^\infty{dk\,\frac{J_1(kQ)}{m^2 + k^2}}.
\end{equation} We observe that ${\cal F}(mQ) \rightarrow m/Q$ as $Q
\rightarrow 0$ and ${\cal F}(mQ) \rightarrow 1/Q^2$ as $Q \rightarrow
\infty$. The similarly transformed fields are \begin{equation}
\tilde{A}_l({\bf x}) = A_l({\bf x}) - im\sum_{{\bf
k},\,n=1,2}{\frac{\epsilon_{ln}k_n j_0({\bf k})e^{i{\bf k \cdot x}}}{k^2
\omega^2}}, \label{eq:altilde} \end{equation} \begin{equation}
\tilde{\Pi}_l({\bf x}) = \Pi_l({\bf x}) + \frac{i}{2}\sum_{\bf
k}{\left\{\frac{1}{k^2} + \frac{1}{\omega^2}\right\}k_l j_0({\bf
k})e^{i{\bf k \cdot x}}}, \label{eq:pitilde} \end{equation}
\begin{equation} \tilde{A}_0({\bf x}) = A_0({\bf x}), \end{equation}
\begin{equation} \tilde{G}({\bf x}) = G({\bf x}), \end{equation}
\begin{equation} \tilde{\cal G} = \partial_l \Pi_l -
\frac{1}{4}m\epsilon_{ln}F_{ln}, \label{eq:gtilde} \end{equation}
\begin{equation} \tilde{\psi}({\bf x}) = \psi({\bf x})e^{{\cal D}({\bf
x})} \label{eq:psitilde} \end{equation} where \widetext \begin{equation}
{\cal D}({\bf x})=\frac{e}{2}\int{d{\bf y}\,\left[\xi({\bf |x -
y|})\frac{\partial}{\partial y_l} A_l({\bf y}) + \eta({\bf | x -
y|})\epsilon_{ln}\frac{\partial}{\partial y_l}\Pi_n({\bf y})\right]},
\end{equation} \narrowtext \noindent and \begin{equation}
\tilde{E}_l({\bf x}) = E_l({\bf x}) + {\cal E}_l({\bf x}),
\label{eq:electrictilde} \end{equation} \begin{equation} \tilde{B}({\bf
x}) = B({\bf x}) + {\cal B}({\bf x}), \label{eq:magnetictilde}
\end{equation} where \begin{equation} {\cal E}_l({\bf x}) = -
\frac{1}{2\pi}\frac{\partial}{\partial x_l}\int{d{\bf y}\,K_0(m{\bf | x
- y|})j_0({\bf y})}, \end{equation} and \begin{equation} {\cal B}({\bf
x}) = -\frac{m}{2\pi}\int{d{\bf y}\,K_0(m{\bf | x - y|})j_0({\bf y})}.
\end{equation} \noindent In this equivalent, alternative,
representation, $\exp(-i\tilde{H}t)$ is the time-translation operator. A
time-translation operator will time-translate state vectors entirely
within the physical subspace in the transformed representation, if it is
entirely devoid of $a_R^\star({\bf k})$ and $a_R({\bf k})$ operators, or
if it contains them at most in the combination $[a_R^\star({\bf
k})a_Q({\bf k}) + a_Q^\star({\bf k})a_R({\bf k})]$. Inspection of
Eq.~(\ref{eq:HItilde}) confirms that $\tilde{H}$ is, in fact, entirely
devoid of $a_R^\star({\bf k})$ and $a_R({\bf k})$ operators, so that the
time-translation operator, $\exp(-i\tilde{H}t)$, correctly satisfies
this requirement. Observable states in the alternative transformed
representation are described by state vectors in $\{|n\rangle\}$ which
we designate as $|N\rangle$. These states consist of massive photons,
electrons and positrons only, and have a positive norm. The operator
$\exp(-i\tilde{H}t)$ time-translates such state vectors by generating a
new state vector, at a later time $t$, which consists of further
positive-norm state vectors $|N^\prime\rangle$, as well as additional gh
ost states, all of which are represented by products of $a_Q^\star({\bf
k})$ operators acting on positive norm observable states. At all times,
the positive norm states alone just saturate unitarity. We can define a
quotient space, which is the residue of $\{|n\rangle\}$ after all zero-
norm states have been excised from it. We can also define another
Hamiltonian, $\tilde{H}_{quot}$, which consists of those parts of
$\tilde{H}$ that remain after we have removed all the terms in which
$a_Q^\star({\bf k})$ or $a_Q({\bf k})$ is a factor; $\tilde{H}_{quot}$
is given by \begin{eqnarray} \tilde{H}_{quot} &=& \sum_{\bf k}{
\frac{\omega}{2}\left\{a({\bf k})a^{\dagger}({\bf k}) + a^{\dagger}({\bf
k})a({\bf k})\right\}}\nonumber\\ &+& \frac{1}{4\pi}\int{d{\bf x}d{\bf
y}\,j_0({\bf x})j_0({\bf y})K_0(m{\bf |x-y|})}\nonumber\\ &+& \int{d{\bf
x}d{\bf y}\,j_0({\bf x})\epsilon_{ln}j_l({\bf y})({ x-y})_n{\cal
F}(m{\bf |x-y|})}\nonumber\\ &+&\sum_{{\bf k},\,l=1,2}{\frac{mk_{l}}{
\sqrt{2}k \omega^{3/2}}\left\{j_{l}({\bf k})a^{\dagger}({\bf k}) +
j_{l}(-{\bf k})a({\bf k})\right\}}\nonumber\\ &+& \sum_{{\bf
k},\,l,n=1,2}{\frac{i\epsilon_{ln}k_{n}}{k\sqrt{2\omega}}
\left\{j_{l}({\bf k})a^{\dagger}({\bf k}) - j_{l}(-{\bf k})a({\bf
k})\right\}}\nonumber\\ &+& H_{e\bar{e}}. \end{eqnarray} \noindent It is
manifest that $\exp(-i\tilde{H}t)|N\rangle$ and $\exp(-
i\tilde{H}_{quot}t)|N\rangle$, where $|N\rangle$ is in the positive-norm
quotient space, have identical projections on other state vectors in the
quotient space. The parts of $\tilde{H}$ that contain $a_Q^\star({\bf
k})$ or $a_Q({\bf k})$ as factors therefore do not play any role in the
time evolution of state vectors within the quotient space of observable
states, and can not have any effect on the physical predictions of the
theory.

We observe that $\tilde{H}_I$ describes the interaction of massive
photons with charged currents. It also describes non-local interactions
between charged fermions. These interactions include the $2+1$
dimensional analog of the Coulomb interaction, with the inverse power of
the distance between charges replaced by the modified Bessel function
$K_0(m|{\bf x-y}|)$. Another such interaction, which has no analog in
$3+1$ dimensional QED, couples charges and the transverse components of
currents. The expressions for the non-local interactions among charge
and current densities that result from these ``ghost'' field components
are well-behaved and free from the kind of infrared singularities that
one might anticipate from massless particle exchange in a $2+1$
dimensional model. The non-local interaction between charge and current
densities, which is without an analog in $3+1$ dimensional QED, is a
manifestation of an important difference between these two models. In
$3+1$ dimensional QED, the magnetic field and the transverse electric
field consist entirely of propagating photons (massless, in that case).
The longitudinal electric field contains no propagating photons at all.
In $2+1$ dimensional QED with a topological mass, however, propagating
photon modes occur in the longitudinal as well as in the transverse
electric field; and ${\cal B}({\bf x})$ is a part of the magnetic field
which is a non-local integral over the charge density, just as is the
corresponding longitudinal electric field ${\cal E}_l({\bf x})$. ${\cal
E}_l({\bf x})$ and ${\cal B}({\bf x})$ accompany charges rigidly as they
move in the plane. ${\cal E}_l({\bf x})$ behaves very much like the
Coulomb field in $3+1$ dimensional QED; ${\cal B}({\bf x})$ follows the
charge like the projection of an extended flux whisker on the plane.
Although ${\cal E}_l({\bf x})$ and ${\cal B}({\bf x})$ are fully
determined by Gauss's Law, the electric and magnetic fields, in their
entirety, are not. Gauss's Law does not force any specific relationship
between the charge density and the field strengths. It is possible, for
example, to form state vectors with electron-photon combinations in
which expectation values of $B({\bf x})$, together with the ${\cal
B}({\bf x})$ accompanying a given charge distribution, can produce
practically any value of $\tilde{{\cal B}}({\bf x})$, the total magnetic
field in the transformed representation. These state vectors, when
described in the alternate transformed representation, obey
$\tilde{\Omega}({\bf k})|n\rangle = a_Q({\bf k})|n\rangle = 0$ and are
entirely consistent with Gauss's Law.

The electron field operator, $\psi$, can be expanded as superpositions
of electron and positron creation and annihilation operators, and 2-
component spinors that constitute the single-particle solutions of the
Dirac Equation. In the transformed representation, $e^\dagger({\bf
p})|0\rangle$ represents the one-electron state which obeys Gauss's Law,
since $a_Q({\bf k})e^\dagger({\bf p})|0\rangle = 0$. Similarly,
$\psi({\bf x})$ is a gauge-invariant spinor field in the transformed
representation, and projects the positron state which obeys Gauss's Law
($\psi^\dagger({\bf x})$ projects the corresponding electron state) out
of the vacuum $|0\rangle$. The corresponding gauge-invariant field, in
the original representation, is $\psi({\bf x})\exp(-{\cal D}({\bf x}))$.
To demonstrate the gauge-invariance of this spinor operator, we note
that gauge transformations are implemented by $\zeta \rightarrow \exp(-
{\cal T})\zeta\exp({\cal T})$, where ${\cal T} = i\int{d{\bf x}\,{\cal
G}({\bf x})\chi({\bf x})}$ and $\chi({\bf x})$ is a c-number function.
Combining \begin{equation} e^{-{\cal T}}\psi({\bf x})e^{-{\cal D}({\bf
x})}e^{{\cal T}}= [e^{-{\cal T}}\psi({\bf x})e^{{\cal T}}][e^{-{\cal
T}}e^{-{\cal D}({\bf x})}e^{{\cal T}}], \end{equation} with
\begin{equation} e^{-{\cal T}}\psi({\bf x})e^{{\cal T}} = \psi({\bf
x})e^{-ie\chi}, \end{equation} and \begin{equation} e^{-{\cal T}}e^{-
{\cal D}({\bf x})}e^{{\cal T}} = e^{-({\cal D}({\bf x}) - ie\chi)},
\end{equation} we obtain the result that $\psi({\bf x})\exp(-{\cal
D}({\bf x}))$ is gauge-inva\-riant. In the transformed representation,
gauge transformations are implemented by $\tilde{\cal T}$ instead of
${\cal T}$, and $\tilde{\cal T}$ is given by $\tilde{\cal T} =
i\int{d{\bf x}\,\tilde{\cal G}({\bf x})\chi({\bf x})}$. Since
$\tilde{\cal G}({\bf x})$ commutes with $\psi({\bf x})$, $\psi({\bf x})$
is the form of the spinor field that is gauge-invariant in the
transformed representation.

Finally, we note that the subspace $\{|n\rangle\}$ of the Fock space
$\{|h_i\rangle\}$ appears in two different and distinct contexts in this
theory. In the first context, when used in conjunction with the field
operators $A_l$, $\Pi_l$, ${\cal G}$ and $\psi$, and with the
Hamiltonian $H = H_0+H_I$, it is a space of states for which the
necessary constraints have not been implemented, except in the
physically uninteresting case of non-interacting photons and electrons,
i.e., the case in which $H_I$ has been eliminated from $H$. In this
context, the states $|N\rangle$ in $\{|n\rangle\}$ that represent
physically observable particles, are generally used to designate
incident and scattered states in perturbative S-matrix elements. But,
since these states do not implement Gauss's Law and the gauge condition,
further arguments are necessary to justify their use. Such arguments
will be reviewed in the next section.

In the second context, the subspace $\{|n\rangle\}$ is the space into
which the unitary transformation $U$ maps $\{|\nu\rangle\}$. In this
context, $\{|n\rangle\}$ is used in conjunction with the unitarily
transformed field operators $\tilde{A}_l$, $\tilde{\Pi}_l$, $\tilde{\cal
G}$, and $\tilde{\psi}_l$, and the Hamiltonian $\tilde{H}$. When used in
this context,  $\{|n\rangle\}$ {\it does\/} implement Gauss's Law and
the gauge condition, even in the presence of all electron-photon
interactions. As it happens, the space  $\{|n\rangle\}$, into which
$\{|\nu\rangle\}$ is mapped, coincides with the space which is used
perturbatively with the untransformed version of this theory, and which,
in that context, implements Gauss's Law {\it only in the interaction-
free limit\/} in which $H_I$ has been eliminated from $H$. It is
important to distinguish these two contexts in which  $\{|n\rangle\}$
appears, and to interpret the states consistently with their proper role
in these two distinct cases, i.e., the original, untransformed, and the
transformed versions of this model.

\section{PERTURBATIVE THEORY}

The perturbative S-matrix is given by $S_{f,i}=\delta_{fi}-2\pi i
\delta(E_f - E_i)T_{f,i}$, where $T_{f,i}$ is the transition amplitude
\begin{equation} T_{f,i} = \langle f|H_I + H_I(E_i - H +
i\varepsilon)^{-1}H_I|i\rangle. \label{eq:1pertheory} \end{equation} The
incident state, $|i\rangle$, and the final, scattered state,
$|f\rangle$, are in $\{|n\rangle\}$, and when used in conjunction with
these operators, these states fail to implement Gauss's Law properly. In
order to implement Gauss's Law as well as the gauge condition, we would
have to use the transition amplitude \begin{equation} \bar{T}_{f,i} =
\langle f|\hat{H}_I + \hat{H}_I(E_i-\tilde{H}+i\varepsilon)^{-
1}\hat{H}_I|i\rangle, \end{equation} where $\hat{H}_I = \tilde{H} -
H_0$. In previous work\cite{temp1}, it has been shown that $T_{f,i}$ and
$\bar{T}_{f,i}$ are related by \begin{equation} \bar{T}_{f,i} = T_{f,i}
+ (E_f - E_i){\cal T}^\alpha_{f,i} + i\varepsilon{\cal T}^\beta_{f,i}
\end{equation} where ${\cal T}^\alpha$ and ${\cal T}^\beta$ are matrix
elements previously given\cite{temp1}. Since $E_i=E_f$ in S-matrix
elements, ${\cal T}^\alpha_{f,i}$ and ${\cal T}^\beta_{f,i}$ do not
contribute to the latter unless ${\cal T}^\alpha_{f,i}$ or ${\cal
T}^\beta_{f,i}$ exhibit $(E_i-E_f)^{-1}$ or $(i\varepsilon)^ {-1}$
singularities, respectively. Such singularities can develop only in
self-energy corrections to external lines, and the resulting
contributions are absorbed into wave function renormalization
constants.\cite{temp1} This result allowed us to substitute $T_{f,i}$
for $\bar{T}_{f,i}$ without affecting the expressions for renormalized
S-matrix elements.

Eq.~(\ref{eq:1pertheory}) leads to Feynman rules, which include the
propagator \begin{equation} D_{ij}(x[1], x[2]) =
\langle0|\hat{T}\{A_i(x[1]), A_j(x[2])\}|0\rangle, \end{equation} where
$\hat{T}$ indicates time ordering, and where $x[1]$, $x[2]$ refer to two
space-time points. $A_l(x)$ is the interaction-picture operator
\begin{equation} A_l(x) = e^{-iH_0t}A_l({\bf x})e^{iH_0t}.
\end{equation} Eqs.~(\ref{eq:Alcs}) and (\ref{eq:hfreegauge2}) can be
used to evaluate $A_l(x)$, leading to \widetext \begin{eqnarray} A_l (x)
&=& -\frac{m}{\sqrt{2}}\sum_{{\bf k}}{\frac{k_l}{k \omega^{3/2}}\left[
a({\bf k}) e^{-i k_\mu x^\mu} + a^{\dagger}({\bf k})e^{i k_\mu
x^\mu}\right]}\nonumber\\ &+& \frac{i}{\sqrt{2}}\sum_{{\bf
k},\,n=1,2}{\frac{\epsilon_{ln}k_n}{k \sqrt{\omega}}\left[ a({\bf k})
e^{-i k_\mu x^\mu} - a^{\dagger}({\bf k})e^{i k_\mu
x^\mu}\right]}\nonumber\\ &+&\frac{1}{16}\sum_{{\bf k}}{\frac{v^4
k_l}{k^3 \kappa_0^{5/2}}\left[a_R({\bf k}) e^{i{\bf k \cdot x}} +
a_R^{\star}({\bf k})e^{-i{\bf k \cdot x}}\right]}\nonumber\\ &-&
\frac{8\gamma}{\beta} \sum_{{\bf k}}{\frac{k \kappa_0^{3/2}k_l}{v^2
\omega^2}\left[a_Q({\bf k}) e^{i{\bf k \cdot x}} + a_Q^{\star}({\bf
k})e^{-i{\bf k \cdot x}}\right]}\nonumber\\ &+&8im \sum_{{\bf
k},\,n=1,2}{\frac{\epsilon_{ln}k_n k \kappa_0^{5/2}}{v^4
\omega^2}\left[a_Q({\bf k}) e^{i{\bf k \cdot x}} - a_Q^{\star}({\bf
k})e^{-i{\bf k \cdot x}}\right]}\nonumber\\ &-& 4 \gamma t \sum_{{\bf
k}}{\frac{k\sqrt{\kappa_0}k_l}{v^2}\left[a_Q({\bf k}) e^{i{\bf k \cdot
x}} + a_Q^{\star}({\bf k})e^{-i{\bf k \cdot x}}\right]}\nonumber\\ &-&8
i t \sum_{{\bf k}}{\frac{k^3 \kappa_0^{5/2}k_l}{v^4
\omega^2}\left[a_Q({\bf k}) e^{i{\bf k \cdot x}} - a_Q^{\star}({\bf
k})e^{-i{\bf k \cdot x}}\right]}. \label{eq:Altcs} \end{eqnarray}
\narrowtext \noindent $D_{ij}(x[1],x[2])$ then becomes \widetext
\begin{eqnarray} &&D_{ij}(x[1],x[2]) = - \sum_{\bf k}{\frac{\gamma
v^2}{\beta \kappa_0 k^2 \omega^2} k_i k_j e^{i{\bf k}\cdot({\bf x}_1-
{\bf x}_2)}}\nonumber\\ &-& \sum_{\bf k}{\frac{\gamma v^2}{4\kappa_0^2
k^2} k_i k_j e^{i{\bf k}\cdot({\bf x}_1-{\bf x}_2)}\ (t_1 + t_2)} -
\sum_{\bf k}{\frac{i}{2\omega^2}k_i k_j e^{i{\bf k}\cdot({\bf x}_1-{\bf
x}_2)}\ |t_1 - t_2|}\nonumber\\ &+& \sum_{\bf
k}{\frac{im\epsilon_{ij}}{2\omega^2}e^{i{\bf k}\cdot({\bf x}_1-{\bf
x}_2)}\left\{\theta(t_1 - t_2) - \theta(t_2 - t_1)\right\}}\nonumber\\
&-&\sum_{\bf k}{\frac{im\epsilon_{ij}}{2\omega^2}\left\{e^{-ik_\mu(x_1 -
x_2)^\mu}\theta(t_1 - t_2) - e^{ik_\mu(x_1 - x_2)^\mu}\theta(t_2 -
t_1)\right\}}\nonumber\\ &+&\sum_{\bf
k}{\frac{1}{2\omega}\left\{\delta_{ij} - \frac{k_i
k_j}{\omega^2}\right\}\left\{e^{-ik_\mu(x_1 - x_2)^\mu}\theta(t_1 - t_2)
+ e^{ik_\mu(x_1 - x_2)^\mu}\theta(t_2 - t_1)\right\}},
\label{eq:propagator2} \end{eqnarray} \narrowtext \noindent or
equivalently, in covariant notation, \widetext \begin{eqnarray}
D_{ij}(x[1],x[2]) &=& -\frac{\gamma}{(2\pi)^2}\int{d{\bf
k}\,\frac{v^2}{\beta \kappa_0 k^2 \omega^2}\left\{1 + \frac{\beta
\omega^2}{4\kappa_0}(t_1+t_2)\right\} k_i k_j e^{i{\bf k}\cdot({\bf
x}_1-{\bf x}_2)}}\nonumber\\ &+&\frac{1}{(2\pi)^3}\int{\frac{d^3k}{k_\mu
k^\mu-m^2}\,\left\{\delta_{ij} - \frac{k_i k_j}{k_0^2} -
\frac{im\epsilon_{ij}}{k_0}\right\}e^{-ik_\mu (x_1 - x_2)^\mu}}
\label{eq:feynpro} \end{eqnarray} \narrowtext \noindent where the
spurious pole at $k_0 = 0$ is evaluated with the principal value
prescription. We can use this propagator to obtain the lowest order
electron-electron scattering S-matrix element, $e(p_1)$ $+$ $e(p_2)$
$\rightarrow$ $e(p_1^\prime)+ e(p_2^\prime)$. It is given by \widetext
\begin{eqnarray} S^{(2)}_{e-e} &=& \frac{f(p_1, p_1^{\prime}, p_2,
p_2^{\prime})\bar{u}({\bf p}_1^{\prime})\gamma^\mu u({\bf
p}_1)\bar{u}({\bf p}_2^{\prime})\gamma_\mu u({\bf p}_2)}{(p_1 -
p_1^{\prime})_\mu(p_1 - p_1^{\prime})^\mu - m^2}\nonumber\\ &-&
\frac{f(p_1, p_1^{\prime}, p_2, p_2^{\prime})\bar{u}({\bf
p}_2^{\prime})\gamma^\mu u({\bf p}_1)\bar{u}({\bf
p}_1^{\prime})\gamma_\mu u({\bf p}_2)}{(p_1 - p_2^{\prime})_\mu(p_1 -
p_2^{\prime})^\mu - m^2}\nonumber\\ &+&\frac{m\epsilon^{\mu \nu
\alpha}(p_1 - p_1^{\prime})_{\alpha}f(p_1, p_1^{\prime}, p_2,
p_2^{\prime}) \bar{u}({\bf p}_1^{\prime})\gamma_\mu u({\bf
p}_1)\bar{u}({\bf p}_2^{\prime})\gamma_\nu u({\bf p}_2)}{(p_1 -
p_1^{\prime})_\mu(p_1 - p_1^{\prime})^\mu\left\{(p_1 - p_1^{\prime})_\nu
(p_1 - p_1^{\prime})^\nu - m^2\right\}}\nonumber\\ &-
&\frac{m\epsilon^{\mu \nu \alpha}(p_1 - p_2^{\prime})_{\alpha}f(p_1,
p_1^{\prime}, p_2, p_2^{\prime}) \bar{u}({\bf p}_2^{\prime})\gamma_\mu
u({\bf p}_1)\bar{u}({\bf p}_1^{\prime})\gamma_\nu u({\bf p}_2)}{(p_1 -
p_2^{\prime})_\mu (p_1 - p_2^{\prime})^\mu\left\{(p_1 -
p_2^{\prime})_\nu (p_1 - p_2^{\prime})^\nu - m^2\right\}} \label{eq:see}
\end{eqnarray} \narrowtext \noindent where \widetext \begin{equation}
f(p_1, p_1^{\prime}, p_2, p_2^{\prime}) = \frac{m_e (-ie)^2
(2\pi)^3}{\sqrt{w(p_1)w(p_2)w(p_1^{\prime})w(p_2^{\prime})}}
\delta^{(3)}(p_1^{\prime} + p_2^{\prime} - p_1 - p_2). \end{equation}
\narrowtext

\section{ROTATION OPERATOR FOR CHARGED\\ PARTICLES}

It has been demonstrated that in pure CS theory, a charged particle of
charge $e$ interacting with a CS field in the absence of the Maxwell
kinetic energy, acquires the phase $e^2/m$ when it is rotated through
$2\pi$ radians. Similarly, a state consisting of two charged particles
acquires this phase, when the two particles are interchanged in a $\pi$
radian rotation.\cite{ch,ms,gwsandsodano} This arbitrary phase has been
interpreted as a signal that charged particles, ``dressed'' in the CS
fields required to obey Gauss's Law, manifest anyonic behavior. We will
here examine the effect of $2\pi$ rotations on charged particles in the
same model, but with a Maxwell term included---i.e., $2+1$ dimensional
QED with a topological mass, interacting with electrons described by the
$2+1$ dimensional Dirac Equation. We will argue that, in this case,
charged particle states only change signs under a $2\pi$ rotation, and
continue to behave like fermions even when they carry the fields they
need to obey Gauss's Law.

The canonical (Noether) angular momentum of the model is
\begin{equation} J = J_g + J_e, \end{equation} where $J_g$ and $J_e$ are
the angular momenta of the gauge field and the spinors, respectively.
$J_g$ and $J_e$ are given by \widetext \begin{equation} J_g = -
\int{d{\bf x}\,\Pi_i x_l\epsilon_{ln}\partial_nA_i} + \int{d{\bf x}\,G
x_l\epsilon_{ln}\partial_nA_0}- \int{d{\bf x}\,\epsilon_{ln}\Pi_l A_n}
\label{eq:jg} \end{equation} \narrowtext \noindent and \begin{equation}
J_e = -i\int{d{\bf x}\,\psi^\dagger x_l\epsilon_{ln}\partial_n\psi} -
\int{d{\bf x}\,\psi^\dagger \frac{\gamma_0}{2}\psi}. \label{eq:je}
\end{equation} In each of these expressions, the last term refers to the
spin, the others to the orbital angular momentum. Direct calculation
verifies that $[H, J] = 0$, so that the total angular momentum is time
independent.

The interpretation of these angular momentum operators, in terms of the
angular momenta of the constituent particle mode excitations, is greatly
simplified when we replace single particle plane waves with eigenstates
of angular momentum. We therefore substitute gauge field annihilation
and creation operators describing excitations with definite angular
momentum, $\alpha_n(k)$ and $\alpha^\dagger_n(k)$, respectively, for the
corresponding plane wave excitations $a({\bf k})$ and $a^\dagger({\bf
k})$. For this purporse, we use \begin{equation} \alpha_n(k) =
\frac{e^{in\pi/2}}{2\pi}\int{d\tau\,a({\bf k})e^{-in\tau}},
\label{eq:Anan} \end{equation} where $\tau$ is the angle that fixes the
direction of ${\bf k}$ in the plane, and a corresponding expression for
the hermitian adjoints $\alpha^\dagger_n(k)$ and $a^\dagger({\bf k})$.
Similar expressions apply to the ghost operators $\alpha_{Q,n}(k)$,
$\alpha_{R,n}(k)$, and their respective adjoints,
$\alpha_{Q,n}^\star(k)$, $\alpha_{R,n}^\star(k)$. In all cases,
Eq.~(\ref{eq:Anan}) relates these operators with their respective plane
wave analogs, $a_Q({\bf k})$, $a_R({\bf k})$, $a_Q^\star({\bf k})$ and
$a_R^\star({\bf k})$. The gauge fields $A_i({\bf x})$, $\Pi_i({\bf x})$,
$A_0({\bf x})$, and $G({\bf x})$ are given, in terms of $\alpha_{n}(k)$,
$\alpha_{n}(k)$, $\alpha_{Q,n}(k)$, $\alpha_{R,n}(k)$ and their
adjoints, as follows: \widetext \begin{eqnarray} &&A_l(\rho,\varphi) =
\frac{i}{2}\sum_{n}{\int{\frac{kdk}{2\pi}}\frac{1}{\sqrt{2\omega}}
\left\{[\Xi_{l,n}^{(+)} - \frac{m}{\omega}\Xi_{l,n}^{(-)}]\alpha_n(k)-
[\Xi_{l,n}^{(+)\,*} - \frac{m}{\omega} \Xi_{l,n}^{(-
)\,*}]\alpha^\dagger_n(k)\right\}}\nonumber\\ &+& \
4i\sum_{n}{\int{\frac{kdk}{2\pi}}\frac{k^2\kappa_0^{5/2}m}{v^4\omega^2}
\left\{[\Xi_{l,n}^{(+)} - \frac{\gamma v^2}{\beta \kappa_0
m}\Xi_{l,n}^{(-)}]\alpha_{Q,n}(k) - [\Xi_{l,n}^{(+)\,*} - \frac{\gamma
v^2}{\beta \kappa_0 m} \Xi_{l,n}^{(-)\,*}]
\alpha^\star_{Q,n}(k)\right\}}\nonumber\\ &&{\ \ \ \ \ \ \ \ \ \ \ \ \ }
+
\frac{i}{32}\sum_{n}{\int{\frac{kdk}{2\pi}}\frac{v^4}{k^2\kappa_0^{5/2}}
\left\{\Xi_{l,n}^{(-)}\alpha_{R,n}(k) - \Xi_{l,n}^{(-
)\,*}\alpha_{R,n}^\star(k)\right\}}, \label{eq:Alcsam} \end{eqnarray}
\narrowtext \noindent where $\Xi^\pm_{l,n}$ are solutions of the two-
dimensional Laplace equation in cylindrical coordinates given by
\begin{equation} \Xi_{l,n}^{(\pm)} =
\epsilon_l^{(+)}J_{n+1}(k\rho)e^{i(n+1)\varphi} \pm \epsilon_l^{(-
)}J_{n-1}(k\rho)e^{i(n-1)\varphi}, \end{equation} with \begin{equation}
\epsilon^{(\pm)}_l = \pm i \epsilon_{l,1} + \epsilon_{l,2},
\end{equation} \widetext \begin{eqnarray} \Pi_l ({\bf x}) &=&
\frac{1}{2}\sum_{n}{\int{\frac{kdk}{2\pi}}\frac{m}{4\sqrt{2\omega}}
\left\{[\frac{2\omega}{m}(1-\frac{m^2}{2\omega^2})\Xi_{l,n}^{(+)} -
\Xi_{l,n}^{(-)}]\alpha_n(k)\right.}\nonumber\\ &&{\ \ \ \ }+\
\left.[\frac{2\omega}{m}(1-\frac{m^2}{2\omega^2})\Xi_{l,n}^{(+)\,*} -
\Xi_{l,n}^{(-)\,*}]\alpha^\dagger_n(k)\right\}\nonumber\\ &-&\
\sum_{n}{\int{\frac{kdk}{2\pi}}\frac{2mk^2\kappa_0^{3/2}}{v^2\omega^2}
\left\{[\frac{\gamma}{\beta}\Xi_{l,n}^{(+)} + \frac{m\kappa}{v^2} (1-
\frac{m^2}{2\omega^2}) \Xi_{l,n}^{(-
)}]\alpha_{Q,n}(k)\right.}\nonumber\\ &&{\ \ \ \ }+\
\left.[\frac{\gamma}{\beta}\Xi_{l,n}^{(+)\,*} + \frac{m\kappa}{v^2} (1-
\frac{m^2}{2\omega^2}) \Xi_{l,n}^{(-
)\,*}]\alpha_{Q,n}^\star(k)\right\}\nonumber\\ &+&
\frac{1}{64}\sum_{n}{\int{\frac{kdk}{2\pi}}\frac{mv^4}{k^2\kappa_0^{5/2}
} \left\{\Xi_{l,n}^{(+)}\alpha_{R,n}(k) +
\Xi_{l,n}^{(+)\,*}\alpha_{R,n}^\star(k)\right\}}, \label{eq:Plcsam}
\end{eqnarray} \begin{equation} A_0(\rho,\varphi) = -4i \gamma
\sum_{n}{\int{\frac{kdk}{2\pi}}\frac{k\sqrt{\kappa_0}}{v^2}J_n(k\rho)
\left[\alpha_{Q,n}(k) e^{in\varphi} - \alpha^\star_{Q,n}(k) e^{-
in\varphi}\right]}, \label{eq:A0csam} \end{equation} and
\begin{eqnarray} G(\rho,\varphi) &=& \frac{1}{8\gamma}
\sum_{n}{\int{\frac{kdk}{2\pi}}\frac{v^2}{k\sqrt{\kappa_0}}
J_n(k\rho)\left[\alpha_{R,n}(k) e^{in\varphi} + \alpha^\star_{R,n}(k)
e^{-in\varphi}\right]}\nonumber\\ &+& \frac{16}{\beta}
\sum_{n}{\int{\frac{kdk}{2\pi}}\frac{k^3 \kappa_0^{7/2}}{v^4
\omega^2}J_n(k\rho)\left[\alpha_{Q,n}(k) e^{in\varphi} +
\alpha^\star_{Q,n}(k) e^{-in\varphi}\right]}. \label{eq:Gcsam}
\end{eqnarray} \narrowtext \noindent The operators $\alpha_i(k)$ an
d $\alpha_j^\dagger(k)$ obey the commutation rules \begin{equation}
[\alpha_i(k), \alpha_j^\dagger(q)] = \frac{2\pi}{k}\delta(k-
q)\delta_{ij}, \end{equation} and \FL \begin{equation} [\alpha_{Q,i}(k),
\alpha_{R,j}^\star(q)] = [\alpha_{R,i}(k), \alpha_{Q,j}^\star(q)] =
\frac{2\pi}{k}\delta(k-q)\delta_{ij}. \end{equation} We also define the
single particle solutions of the Dirac Equation, \widetext
\begin{equation} u_{+}(n,k; \rho, \varphi) =
\frac{1}{\sqrt{2\bar{\omega}_k(M+\bar{\omega}_k)}}\left(\begin{array}{c}
ikJ_{n}(k\rho)e^{in\varphi} \\
(M+\bar{\omega}_k)J_{n+1}(k\rho)e^{i(n+1)\varphi}\end{array}\right)
\end{equation} and \begin{equation} u_{-}(n,k; \rho, \varphi) =
\frac{1}{\sqrt{2\bar{\omega}_k(M+\bar{\omega}_k)}}\left(\begin{array}{c}
(M+\bar{\omega}_k)J_{n}(k\rho)e^{in\varphi} \\
ikJ_{n+1}(k\rho)e^{i(n+1)\varphi}\end{array}\right) \end{equation}
\narrowtext \noindent where $J_s(x)$ is the Bessel function and
$\bar{\omega}_k = M^2 + k^2$. The $u_\pm(n,k;\rho,\varphi)$ are
normalized so that \begin{equation} \int{\frac{\rho d\rho
d\varphi}{(2\pi)^2}\,u_\pm^\dagger(n,k;\rho,\varphi)
u_\pm(n,k;\rho,\varphi)} = 1, \end{equation} and \begin{equation}
\int{\frac{\rho d\rho
d\varphi}{(2\pi)^2}\,u_\pm^\dagger(n,k;\rho,\varphi)
u_\mp(n,k;\rho,\varphi)} = 0, \end{equation} and they obey the equation
\FL \begin{equation} -(i\epsilon_{ln}x_l\partial_n +
\frac{\gamma^0}{2})u_\pm(n,k; \rho, \varphi) = (n+\frac{1}{2})u_\pm(n,k;
\rho, \varphi). \end{equation} We expand the spinor field $\psi({\bf
x})$ in terms of these angular momentum eigenstates and obtain \widetext
\begin{equation} \psi(\rho, \varphi) =
\sum_{n}{\int{\frac{kdk}{2\pi}\,\left\{b_n(k)u_{+}(n,k;\rho,\varphi) +
\bar{b}_n^\dagger(k)u_{-}(n,k;\rho,\varphi)\right\}}} \label{eq:psiam}
\end{equation} \narrowtext \noindent where $b_n(k)$ and $\bar{b}_n(k)$
are the electron and positron annihilation operators, respectively, for
states with definite angular momentum; $b_n^\dagger(k)$ and
$\bar{b}_n^\dagger(k)$ are the corresponding creation operators. When
the corresponding expressions for the gauge fields $A_i({\bf x})$,
$\Pi_i({\bf x})$, $A_0({\bf x})$, and $G({\bf x})$ together with
Eq.~(\ref{eq:psiam}) are used in expanding the angular momentum
operator, we obtain \widetext \begin{equation} J_g =
\sum_n{\int{\frac{kdk}{2\pi}\, n\left[\alpha_n^\dagger(k)\alpha_n(k) +
\alpha_{R,n}^\star(k)\alpha_{Q,n}(k) +
\alpha_{Q,n}^\star(k)\alpha_{R,n}(k)\right]}} \label{eq:jgam}
\end{equation} \narrowtext \noindent and \begin{equation} J_e =
\sum_n{\int{\frac{kdk}{2\pi}\,(n+\frac{1}{2})\left[b_n^\dagger(k)b_n(k)
- \bar{b}_n^\dagger(k)\bar{b}_n(k)\right]}}. \label{eq:jeam}
\end{equation} Questions have been raised about whether the canonical
angular momentum operator is appropriate for determining the phase
acquired by a charged particle under a $2\pi$
rotation.\cite{gwsandsodano,gwsandwije}  We will therefore discuss the
properties of $J$ in some detail. We make the following observations:

\medskip 1) Direct calculation verifies that $[H,J] = 0$, so that the
total angular momentum is time-independent.

\medskip 2) $J$ is invariant to a time-independent gauge transformation,
$\psi \rightarrow \exp(ie\chi)\psi$, $A_l \rightarrow A_l +
\partial_l\chi$ and $\Pi_l \rightarrow \Pi_l +
(1/2)m\epsilon_{ln}\partial_n\chi$. Under such a gauge transformation,
\begin{equation} J \rightarrow J + \int{d{\bf x}\,{\cal G}({\bf
x})\epsilon_{ln}x_l\partial_n\chi({\bf x})}; \end{equation} and since
matrix elements $\langle \nu^\prime | {\cal G}({\bf x}) |\nu\rangle$
vanish for $|\nu\rangle$ and $|\nu^\prime\rangle$ in $\{|\nu\rangle\}$,
$J$ remains untransformed within the physical subspace. As we have shown
in Sec. III, all state vectors for observable systems always remain in
the physical subspace $\{|\nu\rangle\}$, since the time evolution
operator cannot transport them out of it. Therefore, the only
significance we can consistently attach to a dynamical variable in this
theory resides in its matrix elements within this physical subspace. For
the canonical angular momentum, $J$, these matrix elements are totally
unaffected by time-independent gauge transformations. Time-depen\-dent
gauge transformations cannot be implemented with\-in this formulation,
because a time-depen\-dent $\chi$ function is not consistent with a
temporal gauge formulation. The canonical angular momentum is similar to
the Hamiltonian and to the linear momentum in the following respects:
Its functional form is specific to a formulation in a particular gauge.
Its matrix elements in the physical subspace are invariant to gauge
transformations permitted within a particular gauge (e.g., it is limited
to time-independent $\chi$ functions in the temporal gauge). And its
validity as a dynamical variable is completely compatible with these
properties.

\medskip 3) The eigenvalues of $J$ are integral for a photon state, and
half-integral for an electron or positron state. This follows
immediately from Eqs.~(\ref{eq:jgam}) and (\ref{eq:jeam}).

\medskip 4) The rotation operator, $R(\theta) = \exp(iJ\theta)$, rotates
particles states correctly. The electron state $|N\rangle =
b^\dagger_N|0\rangle$ is rotated into $|N^\prime\rangle =
\exp[i(N+1/2)\theta]|N\rangle$. To observe the effect of this rotation
on the field $\psi({\bf x})$, we consider the matrix element of
$\psi({\bf x})$ between the vacuum and a one-electron state,
$\langle\psi_N({\bf x})\rangle = \langle 0 |\psi({\bf
x})b^\dagger_N(p)|0\rangle$. When we rotate the state $|N\rangle$ into
$|N^\prime\rangle$, we obtain $\langle\psi_{N^\prime(\theta)}({\bf
x})\rangle = \langle 0 |\psi({\bf x})R(\theta)b^\dagger_N(p)|0\rangle$;
since $\langle\psi_N({\bf x})\rangle = u_{+}(N,p;\rho,\varphi)$,
\begin{equation} \langle\psi_{N^\prime(\theta)}({\bf x})\rangle =
e^{i(N+1/2)\theta} u_{+}(N,p;\rho,\varphi). \end{equation}
$\langle\psi_{N^\prime(\theta)}({\bf x})\rangle$ can therefore be
expressed as \widetext \begin{equation}
\langle\psi_{N^\prime(\theta)}({\bf x})\rangle =
\frac{1}{\sqrt{2\bar{\omega}_k(M +
\bar{\omega}_k)}}\left(\begin{array}{c}
e^{i\theta/2}ipJ_N(kr)e^{iN(\varphi + \theta)}\\ e^{-i\theta/2}(M +
\bar{\omega}_k)J_{N+1}(kr)e^{i(N+1)(\varphi + \theta)}
\end{array}\right) \end{equation} \narrowtext \noindent and,
equivalently, as \begin{equation} \langle\psi_{N^\prime(\theta)}({\bf
x})\rangle = e^{-
i\gamma^1\gamma^2(\theta/2)}u_+(N,k;\rho,\varphi+\theta), \end{equation}
as required. In the case of the gauge field, the matrix element of
$A_l({\bf x})$ between the vacuum and a one-photon state
$\alpha_N^\dagger(k)|0\rangle$ is $\langle[A_l]_N({\bf x})\rangle =
\langle 0 |A_l({\bf x})\alpha_N^\dagger(k)|0\rangle$; after a rotation,
this matrix element is $\langle[A_l]_{N^\prime(\theta)}({\bf x})\rangle
= \langle 0|A_l({\bf x})R(\theta)\alpha_N^\dagger(k)|0\rangle$.
$\langle[A_l]_N({\bf x})\rangle$ is given by \widetext \begin{equation}
\langle[A_1]_N({\bf x})\rangle =
\frac{i}{\sqrt{2k\omega}}\left\{\frac{1}{2}(1 -
\frac{m}{\omega})J_{N+1}(kr)e^{i(N+1)\varphi} + \frac{1}{2}(1 +
\frac{m}{\omega})J_{N-1}(kr)e^{i(N-1)\varphi}\right\} \end{equation} and
\begin{equation} \langle[A_2]_N({\bf x})\rangle =
\frac{1}{\sqrt{2k\omega}}\left\{\frac{1}{2}(1 -
\frac{m}{\omega})J_{N+1}(kr)e^{i(N+1)\varphi} - \frac{1}{2}(1 +
\frac{m}{\omega})J_{N-1}(kr)e^{i(N-1)\varphi}\right\}. \end{equation}
\narrowtext \noindent The rotated matrix elements then are easily shown
to be \FL \begin{equation} \langle[A_1]_{N^\prime(\theta)}({\bf
x})\rangle = \langle[A_1]_N({\bf x})\rangle \cos \theta +
\langle[A_2]_N({\bf x})\rangle\sin\theta, \end{equation} and \FL
\begin{equation} \langle[A_2]_{N^\prime(\theta)}({\bf x})\rangle = -
\langle[A_1]_N({\bf x})\rangle \sin \theta + \langle[A_2]_N({\bf
x})\rangle\cos\theta, \end{equation} again as required. It is an
immediate corollary of the preceding demonstration that, for $\theta =
2\pi$, $\langle[A_l]_{N^\prime(2\pi)}({\bf x})\rangle =
\langle[A_l]_N({\bf x})\rangle$, while
$\langle\psi_{N^\prime(2\pi)}({\bf x})\rangle = -\langle\psi_N({\bf
x})\rangle$. The effect of the rotation operator, $R(\theta)$, therefore
is that the fermion states change sign in a $2\pi$ rotation, but do not
acquire any further arbitrary phase.

Finally, the following crucial question remains to be considered: The
``bare'' one-electron state $|N\rangle$ violates the constraint
$\Omega({\bf k})|N\rangle = 0$, and is not in the physical subspace,
$\{|\nu\rangle\}$. Since it is known that the implementation of Gauss's
Law is responsible for manifestations of anyonic
behavior,\cite{ch,ms,gwsandsodano} it is necessary to go beyond the
examination of how simple perturbative one-particle states rotate in the
plane. To investigate the behavior, under $2\pi$ rotations, of the
electron state which is attached to the electric and magnetic fields
${\cal E}_l({\bf x})$ and ${\cal B}({\bf x})$ respectively, so that it
obeys Gauss's Law, we must substitute $e^{-D}|N\rangle$ for the electron
state $|N\rangle$. Equivalently, we may use the alternate, transformed
formalism. In that case, we can continue to use $|N\rangle$ as the
electron state, and $\psi({\bf x})$ as the gauge-invariant spinor field,
and substitute $\tilde{J} = \exp(-D)J\exp(D)$ for $J$ to represent the
angular momentum. This approach, when applied to the canonical angular
momentum operator in pure CS theory, demonstrates that $\tilde{J}$
differs from $J$, and that the difference accounts for the phase $\eta =
(Q^2/2\pi m)$, where $Q$ is the electron charge \begin{equation} Q =
e\sum_n{\int{\frac{kdk}{2\pi}}\,[b^\dagger_n(k)b_n(k) -
\bar{b}^\dagger_n(k)\bar{b}_n(k)]}. \end{equation} The difference
between $\tilde{J}$ and $J$ accounts for the fact the in pure CS theory,
the ``true'' electron state, for which Gauss's Law is implemented,
acquires the arbitrary phase $\eta$ characteristic of an anyon when it
is rotated through $2\pi$.\cite{ms}

In contrast, in the model we are investigating here, $J$ is unaffected
by the unitary transformation \begin{equation} \tilde{J} = e^{-D}Je^{D},
\end{equation} so that $\tilde{J} = J$. This fact indicates that, once
the Maxwell kinetic energy is included in the Lagrangian, electron
states that obey Gauss's Law do not acquire arbitrary phases in $2\pi$
rotations, and should not be expected to behave like anyons. To
demonstrate this feature of this model, we reexpress $D$, given in
Eq.~(\ref{eq:D}), in terms of the gauge field and electron operators
that designate excitations with definite angular momenta. The resulting
expression is \widetext \begin{displaymath} D = \int{\frac{\rho d\rho
d\varphi}{(2\pi)^2}\sum_n{\int{\frac{kdk}{2\pi}\,\frac{v^4}{16k^3\kappa_
0^{5/2}}J_{-n}(k\rho)\left\{\alpha_{R,n}(k)e^{in\varphi}-
\alpha^\star_{R,n}(k)e^{-in\varphi}\right\}j_0(\rho,\varphi)}}}
\end{displaymath} \begin{equation} +\ \int{\frac{\rho d\rho
d\varphi}{(2\pi)^2}\sum_n{\int{\frac{kdk}{2\pi}\,\frac{8\gamma\kappa_0^{
3/2}k}{\beta v^2 \omega^2}J_{-n}(k\rho)
\left\{\alpha_{Q,n}(k)e^{in\varphi}-\alpha^\star_{Q,n}(k)e^{-
in\varphi}\right\}j_0(\rho,\varphi)}}}. \end{equation} \narrowtext
\noindent We use Eqs.~(\ref{eq:jgam}) and (\ref{eq:jeam}) to evaluate
$[J, D]$ in the first term of the Baker-Hausdorff-Campbell expansion of
$\tilde{J}$, and note that $[J_g, D] = -[J_e, D]$, so that $[J, D]=0$.
All higher order commutators in the expansion of $\exp(-iD)J$ $\exp(iD)$
then vanish trivially, and we have shown that $\tilde{J} = J$.

As a further illustration of this feature of this model, we observe that
$[J,D]$ is not only the first term in the expansion of $\tilde{J}$; it
can also be interpreted as the generator of the rotation group acting on
$D$. To understand $[J,D]$ from this point of view, we refer to
Eq.~(\ref{eq:Dspace}) and note that $D$ is an integral over two spaces
(${\bf x}$ and ${\bf y}$), and that the integrand is a product of two
scalar operators, $\partial_lA_l({\bf x})$ or
$\epsilon_{ln}\partial_l\pi_n({\bf x})$, and $j_0({\bf y})$, connected
by a c-number scalar invariant function of ${\bf |x - y|}$.  Since
$\partial_lA_l$ and $\epsilon_{ln}\partial_l\pi_n$ are scalar operator,
$[J_g, D]$ is given by \widetext \begin{equation} [J_g, D] = \int{d{\bf
x}d{\bf y}\,\left\{[\epsilon_{ij}x_i\partial_j\partial_lA_l({\bf
x})]\xi({\bf |x-y|}) -
[\epsilon_{ij}x_i\partial_j\epsilon_{ln}\partial_l\Pi_n({\bf
x})]\eta({\bf |x-y|})\right\}j_0({\bf y})} \end{equation} and
\begin{equation} [J_e, D] = \int{d{\bf x}d{\bf
y}\,\left\{\partial_lA_l({\bf x})\xi({\bf |x-y|}) -
\epsilon_{ln}\partial_l\Pi_n({\bf x})\eta({\bf |x-
y|})\right\}\epsilon_{ij}y_i\partial_j j_0({\bf y})}. \end{equation}
\narrowtext \noindent Integrations by parts transfer the orbital angular
momenta \begin{displaymath} \epsilon_{ij}x_i(\partial/\partial x_j)
\end{displaymath} and \begin{displaymath}
\epsilon_{ij}y_i(\partial/\partial y_j) \end{displaymath}
from the gauge fields and $j_0$, respectively, to the functions
$\xi({\bf |x-y|})$ and $\eta({\bf |x-y|})$. The result is that the
orbital angular momenta $\epsilon_{ij}x_i(\partial/\partial x_j)$ and
$\epsilon_{ij}y_i(\partial/\partial y_j)$ appear in $[J,D]$ only in the
expressions \begin{displaymath} [\epsilon_{ij}x_i(\partial/\partial x_j)
+ \epsilon_{ij}y_i(\partial/\partial y_j)]\xi({\bf |x-y|})
\end{displaymath} and \begin{displaymath}
[\epsilon_{ij}x_i(\partial/\partial x_j) +
\epsilon_{ij}y_i(\partial/\partial y_j)]\eta({\bf |x-y|}).
\end{displaymath} Moreover, $[\epsilon_{ij}x_i(\partial/\partial x_j) +
\epsilon_{ij}y_i(\partial/\partial y_j)]$ can be expressed as
\begin{displaymath} [\epsilon_{ij}r_i(\partial/\partial
r_j)+\epsilon_{ij}R_i(\partial/\partial R_j)], \end{displaymath} where
${\bf r = x - y}$ and ${\bf R = (x + y)}/2$. \begin{displaymath}
[\epsilon_{ij}r_i(\partial/\partial
r_j)+\epsilon_{ij}R_i(\partial/\partial R_j)]\xi(r) \end{displaymath}
and \begin{displaymath} [\epsilon_{ij}r_i(\partial/\partial
r_j)+\epsilon_{ij}R_i(\partial/\partial R_j)]\eta(r) \end{displaymath}
vanish trivially because neither $\xi(r)$ nor $\eta(r)$ have any angular
dependence in $r$, and both are independent of $R$. The result that
$[J,D]=0$ therefore depends on the fact that $\xi(r)$ and $\eta(r)$ are
independent of ${\bf R}$ and of the angle in ${\bf r}$. There are,
therefore, basic kinematic reasons for the identity of $\tilde{J}$ and
$J$, which apply to this model but not, as Ref. [6] shows, to pure CS
theory.

Lastly, we remark on the anticommutation rules of the electron fields in
this model. The gauge-invariant spinor fields $\psi({\bf x})\exp[-{\cal
D}({\bf x})]$ and $\psi^\dagger({\bf x})\exp[{\cal D}({\bf x})]$, which
pro\-ject the electron and positron states that obey Gauss's Law,  out
of the vacuum, are unitary transforms of $\psi({\bf x})$ and
$\psi^\dagger({\bf x})$, respectively. The anticommutation rules of
$\psi({\bf x})$ $\exp[-{\cal D}({\bf x})]$ and $\psi^\dagger({\bf
x})\exp[{\cal D}({\bf x})]$ are, therefore, identical to those of
$\psi({\bf x})$ and $\psi^\dagger({\bf x})$, and characteristic of
fermions.

In other work on $2+1$ dimensional QED with a topological mass,
Belinfante's ``symmetric'' angular momentum operator, ${\cal J}$, has
been substituted for the canonical angular momentum in determining
whether charged particles acquire arbitrary ``anyonic'' phases when they
are rotated through $2\pi$ radians, or when two charged particles are
exchanged via rotations.\cite{gwsandsodano,gwsandwije} We will briefly
discuss this substitution in the context of our formulation of this
model. ${\cal J}$ is given by \widetext \begin{equation} {\cal J} = -
i\int{d{\bf x}\,\psi^\dagger\epsilon_{ln}x_l(\partial_n - ieA_n)\psi} -
\int{d{\bf x}\,\psi^\dagger\frac{\gamma_0}{2}\psi} -
\frac{1}{2}\int{d{\bf x}\,x_nF_{n0}\epsilon_{ij}F_{ij}}, \end{equation}
\narrowtext \noindent and can be expressed in terms of the canonical
angular momentum, $J$, given in Eq.~(\ref{eq:jg}), as shown by \widetext
\begin{equation} {\cal J} = J + J_{\mbox{\scriptsize surface}} +
J_{\mbox{\scriptsize gauss}} \end{equation} \narrowtext \noindent where
$J_{\mbox{\scriptsize surface}}$ is the surface term \begin{equation}
J_{\mbox{\scriptsize surface}} = \int{d{\bf
x}\,\partial_i(\epsilon_{ln}x_lA_n\Pi_i)}, \end{equation} and where
$J_{\mbox{\scriptsize gauss}}$ is a contribution that vanishes when
${\cal G} = 0$, so that Gauss's Law applies. $J_{\mbox{\scriptsize
gauss}}$ is given by \begin{equation} J_{\mbox{\scriptsize gauss}} =
\int{d{\bf x}\,\epsilon_{ln}x_lA_n{\cal G}}. \end{equation} The argument
has been made that, when Gauss's Law applies, $J_{\mbox{\scriptsize
gauss}}$ vanishes, and ${\cal J}$ then consists only of the canonical
angular momentum, $J$, and the surface term $J_{\mbox{\scriptsize
surface}}$; the latter is then identified with the arbitrary anyonic
phase in the rotation of charged states.\cite{gwsandsodano,gwsandwije}
We will examine here whether $J_{\mbox{\scriptsize gauss}}$ actually
vanishes in the physical subspace $\{|\nu\rangle\}$.

There is no doubt that any operator \begin{equation} W = \int{d{\bf
x}\,U({\bf x}){\cal G}({\bf x})}, \end{equation} where $U({\bf x})$ is a
c-number function, vanishes in $\{|\nu\rangle\}$, and that $W({\bf x},t)
= W({\bf x},t=0)$ in $\{|\nu\rangle\}$, so it remains zero permanently.
However, in $J_{\mbox{\scriptsize gauss}}$, ${\cal G}({\bf x})$ is not
integrated over a c-number function, but over the operator-valued
quantity $\epsilon_{ln}x_lA_n({\bf x})$, so that its behavior under time
evolution is more complicated. To simplify our discussion, we transform
to the alternate representation, as discussed in Sec. III, and obtain
the result that \begin{equation} \tilde{J}_{\mbox{\scriptsize gauss}} =
\int{d{\bf x}\,\epsilon_{ln}x_l\tilde{A}_n\tilde{\cal G}}.
\label{eq:jgtilde} \end{equation} Eqs.~(\ref{eq:altilde}) and
(\ref{eq:gtilde}) allow us to express Eq.~(\ref{eq:jgtilde}) in the form
\widetext \begin{equation} \tilde{J}_{\mbox{\scriptsize gauss}} =
\int{d{\bf x}\,\epsilon_{jl}x_j\left\{A_l({\bf x}) - im\sum_{{\bf
k},n}{\frac{\epsilon_{ln}k_nj_0({\bf k})}{k^2\omega^2}e^{i{\bf k \cdot
x}}}\right\}\tilde{\cal G}({\bf x})} \end{equation} \narrowtext
\noindent with \begin{equation} \tilde{\cal G}({\bf x}) = \sum_{\bf
k}{\frac{8k^3\kappa_0^{5/2}}{v^4}[a_Q({\bf k})e^{i{\bf k \cdot x}} +
a_Q^\star({\bf k})e^{-i{\bf k \cdot x}}]}. \end{equation} We focus
attention on the term \widetext \begin{displaymath} \int{d{\bf
x}\,\epsilon_{ln}x_lA_n({\bf x})\sum_{\bf
k}{\frac{8k^3\kappa_0^{5/2}}{v^4}[a_Q({\bf k})e^{i{\bf k \cdot x}} +
a_Q^\star({\bf k})e^{-i{\bf k \cdot x}}]}} \end{displaymath} \narrowtext
\noindent in $\tilde{J}_{\mbox{\scriptsize gauss}}$, in which the
$a_Q^\star({\bf k})$ and $a_Q({\bf k})$ in $\tilde{\cal G}({\bf x})$
combine with the $a_R^\star({\bf q})$ and $a_R({\bf q})$ in $A_n({\bf
x})$ to contribute the products $a_Q^\star({\bf k})a_R^\star({\bf q})$
and $a_Q({\bf k})a_R({\bf q})$. As we have shown in Sec. III, these
products have the effect of driving state vectors out of the physical
subspace, into the part of the Hilbert Space in which Gauss's Law is no
longer valid. Using $\exp(i{\cal J}\varphi)$ as a rotation operator
would rotate state vectors out of the physical subspace; Gauss's Law,
and the identity that ${\cal J}=J\ +\ $(surface term), would therefore
not apply during and after a rotation performed with the rotation
operator $\exp(i{\cal J}\varphi)$.

Alternatively, if we follow Refs. [9] and [10], and define the physical
subspace as consisting of state vectors $|{\cal N}\rangle$ which obey
${\cal G}({\bf x})|{\cal N}\rangle = 0$, the same difficulty reappears
in a different guise. $\epsilon_{ln}x_lA_n({\bf x})$ and ${\cal G}({\bf
x})$ do not commute, and the operator product $\epsilon_{ln}x_lA_n({\bf
x}){\cal G}({\bf x})$ is ill-defined unless the operator order is
specified carefully. Even if a state $|{\cal N}\rangle$ obeys ${\cal
G}({\bf x})|{\cal N}\rangle = 0$, there will be many states $|{\cal
N}^\prime\rangle$, that arise in the expression $|{\cal
N}^\prime\rangle\langle{\cal N}^{\prime\prime}|\epsilon_{ln}x_lA_n({\bf
x})|{\cal N}\rangle$, where $|{\cal N}^\prime\rangle\langle{\cal
N}^{\prime\prime}|$ is the unit operator in this space, for which ${\cal
G}({\bf x})|{\cal N}^\prime\rangle \neq 0$. The question, whether
$J_{\mbox{\scriptsize gauss}}$ vanishes in the space defined by ${\cal
G}({\bf x})|{\cal N}\rangle = 0$, is therefore dependent on operator
ordering and problematical. In contrast, the ambiguities in the operator
products that appear in the canonical angular momentum are benign, and
are resolved by normal ordering the expressions in Eqs.~(\ref{eq:jgam})
and (\ref{eq:jeam}).

The canonical angular momentum provides us with a satisfactory rotation
operator, which gives completely consistent results for the rotation of
particle states, as well as for matrix elements of field operators; and
it does so when Gauss's Law is implemented for the charged particle
states. It appears to us appropriate, in interpreting this model, not to
arbitrarily abandon the canonical (Noether) angular momentum, and with
it the fermionic property of the electron, when there is no compelling
reason for doing so.

\begin{center} {\bf ACKNOWLEDGMENT} \end{center}

This research was supported by the Department of Energy under Grant No.
DE-AC02-79ER 10336.A.


\begin{references}
\bibitem[15]{comment}We use a standard
representation for the electron field operator in $2+1$ dimensions, such
as the one given in Ref.~[6], but with a choice of gamma matrices given
in the paragraph following Eq.~(\ref{eq:lagrangian2}).
\end{references}
\end{document}